\documentclass[runningheads]{llncs}

\usepackage[T1]{fontenc}
\usepackage{lmodern}
\usepackage{microtype}
\usepackage{booktabs}
\usepackage{tabularx}
\usepackage{array}
\usepackage{xcolor}
\usepackage{tikz}
\usetikzlibrary{arrows.meta,positioning,calc}
\usepackage{graphicx}
\usepackage{float}
\usepackage{url}

\newcolumntype{Y}{>{\raggedright\arraybackslash}X}
\begin{document}

\title{AI-Native Product Management: A Vision}
\titlerunning{AI-Native Product Management: A Vision}
\author{Ricardo Britto\orcidID{0000-0002-7220-9570}}
\authorrunning{Ricardo Britto}
\institute{Ericsson, Sweden}
\maketitle

\begin{abstract}
Artificial intelligence is increasingly used to support product-management tasks. At the same time, AI-native research and development (R\&D) may increase software realization capacity. Organizations may realize software faster than they can identify, qualify, and reconsider what should be realized. This paper asks how product management should be re-envisioned under these conditions. We synthesize research on software product management, product learning, human--AI collaboration, and the Product Management--R\&D boundary. We propose a vision in which AI becomes a persistent participant in product work. The vision covers sensing, exploration, decision-making, realization, and learning. We propose six principles and make the vision tangible through an operating model grounded in human professionals, AI personas, agents, skills, and a shared product context. We further argue that AI-native Product Management and AI-native R\&D can form a continuous product-learning process. Finally, we identify four AI-specific research directions that require empirical investigation.
\keywords{Artificial intelligence \and Software product management \and AI-native R\&D \and Human--AI collaboration \and Product learning}
\end{abstract}

\section{Introduction}
\label{sec:introduction}

Software product management (SPM) connects strategy, customer and market insight, product planning, commercialization, lifecycle management, and software development \cite{vandeweerd2006reference,ebert2007impact,maglyas2017core}. It turns uncertain evidence into decisions about where to invest and what to build, change, or stop. 

Artificial intelligence (AI) is changing SPM. Large language models can synthesize feedback, explore alternatives, analyze product information, and generate early prototypes. They can also process broader market information and support strategic analysis at greater speed and scale. AI can generate and evaluate strategic alternatives, support scenario analysis, and challenge strategic assumptions \cite{csaszar2024}. Nevertheless, most applications in product management still support bounded tasks initiated by a human \cite{witkowski2025ai,moriserra2026llm}.


AI is also changing software development, with AI agents gaining increasing prominence across the entire software lifecycle \cite{hou2024llmse}. AI-native R\&D, in which AI agents and human engineers work side by side to realize software \cite{britto2026ainative}, can increase capacity for software realization. Product management must still identify opportunities, challenge assumptions, choose investments, and revise direction. The opportunity is not only to realize software faster. It is also to explore and stress-test product direction before substantial realization begins.

Our focus is on how AI changes the practice of product management, independent of whether the managed product itself contains AI. We argue that AI should become a persistent participant in product work rather than waiting for individual prompts. Product managers remain responsible for interpreting ambiguous evidence, framing product direction, making consequential trade-offs and commitments, and accepting accountability.




In this paper, we answer the following research question:
\begin{quote}
\emph{RQ: How should product management be re-envisioned when AI becomes a persistent participant in product work?}
\end{quote}

We address this question through a theory-synthesis approach \cite{jaakkola2020conceptual}. The paper makes three contributions:
\begin{enumerate}
    \item \textbf{Conceptualization.} We define AI-native product management around persistent AI participation and distinguish it from task-level assistance.
    \item \textbf{Operating model.} We describe how human judgment, AI personas, agents, skills, shared product context, and AI-native R\&D can interact across product learning.
    \item \textbf{Research agenda.} We identify AI-specific gaps concerning persistent participation, exploration, human--AI allocation and judgment, and the Product Management--R\&D boundary.
\end{enumerate}

The remainder of the paper is organized as follows: Section~\ref{sec:foundations} reviews its foundations. Section~\ref{sec:vision} presents the vision and principles. Section~\ref{sec:operating} makes the vision tangible through an operating model. Section~\ref{sec:illustration} illustrates the vision of a hypothetical use case. Section~\ref{sec:research-agenda} identifies open research directions, and Section~\ref{sec:conclusion} concludes.

\section{Foundations and Related Work}
\label{sec:foundations}

AI-native product management builds on research into SPM as an organizational capability, discovery and experimentation as foundations for learning, and AI support for product work. Research on human--AI collaboration informs how autonomy is allocated, while cross-functional research explains the knowledge boundary between Product Management and R\&D. We bring these streams together to examine what changes when AI participation becomes persistent.

\subsection{Software Product Management as an Organizational Capability}

SPM spans strategy, portfolio management, roadmapping, requirements, releases, commercialization, and lifecycle decisions \cite{vandeweerd2006reference,ebert2007impact,maglyas2017core}. This breadth creates coordination challenges. Product managers work across customers, business functions, development teams, and senior management. Empirical studies report unclear priorities, silos, role ambiguity, and limited authority \cite{springer2022challenges}. 

Agile software development changes how the aforementioned responsibilities are distributed, but it does not remove them \cite{tkalich2022agile}. At scale, product-management configurations also vary with product and development context \cite{moe2024large}.

\subsection{Continuous Discovery, Experimentation, and Product Learning}

Product management requires continuous learning about customer needs, possible solutions, and the value those solutions deliver. Discovery helps identify questions and assumptions worth investigating, while experimentation provides evidence that can guide subsequent product decisions.

Continuous experimentation treats software products as systems for testing assumptions through frequent customer feedback and evidence from use \cite{bosch2012experiments}. HYPEX puts this into practice through feature experiments, where teams develop small parts of a feature and evaluate them with customers to guide further development \cite{olssonbosch2014hypex}. RIGHT takes a broader view of the experimentation system, connecting instrumented product increments and experiment plans with product roadmaps and business strategy \cite{fagerholm2017right}. Both approaches connect development with learning so that new evidence can influence what is built next.

Experimentation is, however, only one source of product knowledge. Decisions about strategy, contracts, architecture, and long-term investments cannot always be assessed through controlled experiments. They also require qualitative insight, technical analysis, and professional judgment. The common principle is that assumptions should be explicit and revisited as evidence changes.

This broader learning process must also look beyond expressed demand. Market orientation research distinguishes between responding to stated needs and discovering latent ones \cite{narver2004responsive}, while market sensing emphasizes continuous learning about customers, competitors, and the wider environment \cite{day1994}. These perspectives complement the distinction between exploring new possibilities and exploiting existing knowledge \cite{march1991exploration}. Prototypes connect exploration with concrete product choices by making ideas and assumptions tangible enough to examine with stakeholders before major commitments \cite{lim2008prototypes,buchenau2000experience}.

\subsection{AI in Product Management}

Research on AI in product management is growing rapidly. Applications include customer analysis, forecasting, requirements work, knowledge extraction, and decision support \cite{witkowski2025ai}. Empirical work also shows how data, AI, and shorter feedback loops are changing strategic digital product management \cite{holmstromolsson2025strategic}.

AI can also support work further upstream. Large language models can generate and evaluate strategic alternatives, scenario analyses, and competitive assessments \cite{csaszar2024}. AI-enabled analytics can also support faster responses to market change \cite{sullivanwamba2024}. These capabilities matter because product management must connect customer evidence with competitors, technology, market developments, and strategic alternatives.

A recent systematic review describes research on large language models in product management as nascent and largely task-focused \cite{moriserra2026llm}. Hence, we do not claim novelty for applying AI to individual activities. The unresolved issue is what changes when sensing, strategic exploration, and other product work become persistent rather than task-triggered.

\subsection{Human--AI Collaboration and AI Autonomy}

Persistent participation does not imply unrestricted autonomy. Automation research shows that information acquisition, analysis, decision selection, and action can be delegated differently \cite{parasuraman2000automation}. Human--AI research also emphasizes appropriate control and productive delegation \cite{amershi2019guidelines,fuegener2022delegation}.

AI-generated representations can also create artificial certainty when detailed outputs make uncertain futures appear more knowable than they are \cite{leonardi2026certainty}. AI-native product management requires sufficient AI initiative to remain genuinely persistent while preserving human judgment for crucial choices.

\subsection{The Product Management--R\&D Boundary in an AI-Native Setting}

Product Management and R\&D contribute different forms of knowledge. Product Management brings customer, market, commercial, and strategic evidence. R\&D brings knowledge of technology, architecture, feasibility, quality, and implementation.

Integrating these forms of knowledge can be difficult. Research on R\&D and market-facing functions describes distinct thought worlds and recurring integration challenges \cite{dougherty1992barriers,griffinhauser1996integration}. Carlile shows that knowledge may need to be transferred, translated, or transformed across functional boundaries \cite{carlile2004boundaries}. Product owners in large-scale agile rely on shared knowledge, shared goals, mutual respect, and high-quality communication with development teams \cite{berntzen2019po}. 

AI-native R\&D increases the need for coordination with Product Management \cite{britto2026ainative}. Faster development can put greater pressure on identifying opportunities and deciding what to build. Persistent AI across both functions may help keep product rationale, technical constraints, uncertainty, and development outcomes connected. Whether this improves mutual understanding remains an open question.

\subsection{Conceptual Synthesis and Gap}

The reviewed work explains product management, product learning, AI-supported tasks, human--AI delegation, and cross-functional knowledge integration. It leaves open how these elements work together when AI participates persistently in product management and across its interaction with R\&D.

We use a theory-synthesis approach to combine these complementary research streams. Table~\ref{tab:vision-derivation} maps the synthesized themes to the proposed vision and their literature foundations. The synthesis provides one structured interpretation, not an exhaustive set of principles or operating elements. The literature supports the underlying concepts; it does not establish the effects of the combined model.

\begin{table}[H]
\caption{Literature foundations of the proposed vision.}
\label{tab:vision-derivation}
\centering
\small
\begin{tabularx}{\textwidth}{@{}>{\raggedright\arraybackslash}p{0.25\textwidth}>{\raggedright\arraybackslash}p{0.32\textwidth}Y@{}}
\toprule
Synthesized theme & Vision element & Literature foundation \\
\midrule
End-to-end product responsibility & Human product judgment; Product Management professional perspectives & SPM scope and organizational capability \cite{vandeweerd2006reference,ebert2007impact,maglyas2017core}. \\
\addlinespace
Continuous evidence-driven exploration and learning & Continuous sensing; exploration; prototyping; challenge; lifecycle learning & Market sensing, exploration, experimentation, and organizational learning \cite{day1994,narver2004responsive,bosch2012experiments,march1991exploration}. \\
\addlinespace
Persistent rather than task-triggered AI participation & Persistent AI participation; expanded strategic exploration & AI in product management and strategic decision-making \cite{witkowski2025ai,moriserra2026llm,csaszar2024}. \\
\addlinespace
Different levels of AI initiative and human judgment & AI-operated, human-directed, and collaborative activities & Human--AI collaboration and delegation \cite{parasuraman2000automation,amershi2019guidelines,fuegener2022delegation}. \\
\addlinespace
Continuity between product and technical reasoning & Product Management--R\&D interleaving; shared semantic context; living product intent & Knowledge boundaries and AI-native R\&D \cite{carlile2004boundaries,berntzen2019po,britto2026ainative}. \\
\bottomrule
\end{tabularx}
\end{table}

\section{A Vision for AI-Native Product Management}
\label{sec:vision}

The foundations above already exist. Our vision combines them by changing the role of AI itself.

\subsection{From AI Assistance to AI-Native Product Management}

Most current uses of AI in product management are assistive. A product manager invokes AI to summarize interviews, draft requirements, analyze feedback, examine a competitor, or create a prototype. AI may improve the task, but the surrounding process remains largely unchanged.

In contrast, in AI-native product management, AI persistently participates in sensing, exploration, decision preparation, product direction, realization, and learning. Humans retain responsibility for product judgment, consequential commitments, and accountability. AI maintains context and can initiate bounded work between human interactions. 

For example, AI may connect new customer evidence with a competitor announcement, a technology trend, and earlier strategic assumptions. It can then surface that the current product direction should be reconsidered without making that decision itself.

Persistent AI also changes the scale of exploration available before commitment. Product professionals can examine competing market interpretations, alternative strategies, scenarios, and emerging threats more frequently \cite{csaszar2024}. AI performs much of the continuous monitoring and synthesis. Humans remain responsible for deciding which signals matter, changing the strategic frame when needed, and choosing the resulting product direction.

We define the six following principles associated with AI-native product management:
\begin{enumerate}
    \item \textbf{Continuous sensing over periodic research.} AI continuously observes customer, competitor, market, product, technology, and operational signals.
    \item \textbf{Exploration over request processing.} AI stress-tests product direction against customer evidence, competitors, emerging trends, scenarios, and alternatives.
    \item \textbf{Prototype before committing.} AI makes ideas cheaper to materialize and assumptions easier to test before major R\&D commitments.
    \item \textbf{Living intent over static artifacts.} AI keeps product direction connected to evidence, assumptions, constraints, and outcomes.
    \item \textbf{Challenge before consensus.} AI searches for counter-evidence, alternative explanations, and competing options.
    \item \textbf{Learn across the lifecycle.} AI connects what was expected with what was realized and observed.
\end{enumerate}

\begin{figure}[t]
\centering
\includegraphics[width=\textwidth]{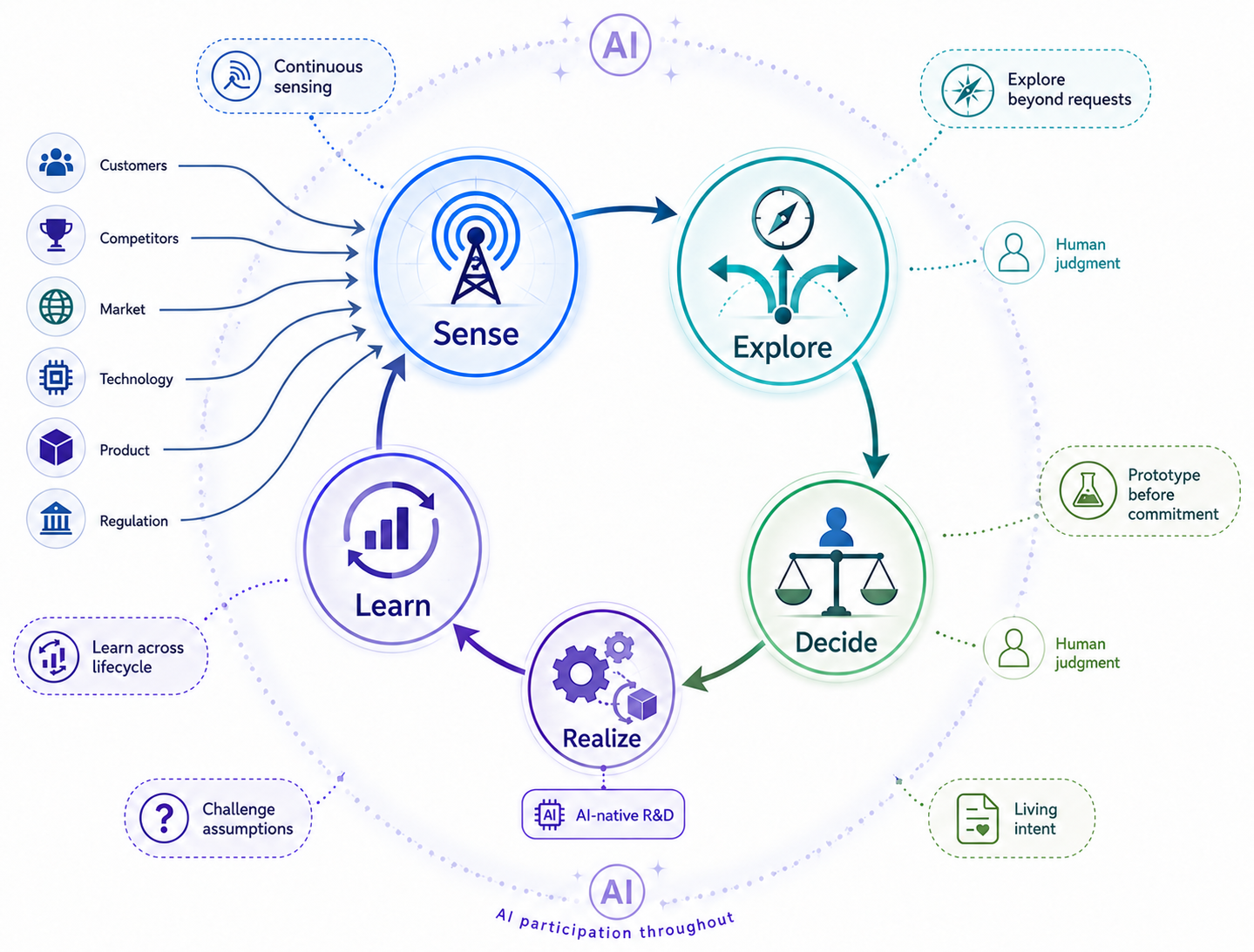}
\caption{The AI-native product learning loop and its underlying principles.}
\label{fig:learning-loop}
\end{figure}

The principles are instantiated through a continuous loop. Figure~\ref{fig:learning-loop} shows AI across the loop rather than as a separate support capability. The loop organizes established product activities, i.e, it is not a new or universal product lifecycle. The proposed change is persistent AI participation across them. The activities overlap, and new evidence can reopen an earlier interpretation or decision.

\textbf{Sense} follows customer, competitor, market, product, technology, business, and operational signals. AI performs much of the monitoring and synthesis while humans decide which changes matter.

\textbf{Explore} turns signals into opportunities and alternative product directions. AI compares competitors, examines trends, generates scenarios, challenges assumptions, and creates prototypes. Humans frame the problem, interpret evidence, and choose which alternatives deserve deeper investigation \cite{narver2004responsive,csaszar2024}.

\textbf{Decide} combines AI-supported analysis with human judgment. AI connects evidence, alternatives, constraints, and uncertainty. Humans remain responsible for strategic trade-offs, investments, and consequential commitments.

\textbf{Realize} connects the chosen product direction with AI-native R\&D \cite{britto2026ainative}. R\&D contributes technical possibilities, constraints, feasibility evidence, and emerging alternatives. These may reopen product exploration when important assumptions change.

\textbf{Learn} connects original expectations with realized and observed outcomes. Customer response, market change, product use, commercial outcomes, and technical evidence become new inputs to sensing and exploration.

Together, the principles shift AI from episodic assistance toward continuous participation. Table~\ref{tab:expected-benefits} summarizes the expected benefits, which remain propositions rather than demonstrated outcomes. The benefits are interdependent. Sensing and prototyping matter only when new evidence can influence product direction.

\begin{table}[H]
\caption{Expected benefits of AI-native product management.}
\label{tab:expected-benefits}
\centering
\small
\begin{tabularx}{\textwidth}{@{}>{\raggedright\arraybackslash}p{0.32\textwidth}Y@{}}
\toprule
Expected benefit & How the principles contribute \\
\midrule
Better sensing and strategic exploration & Connect customer, competitor, market, technology, and product signals and stress-test direction before commitment. \\
Faster learning before commitment & Make important assumptions testable through rapid prototypes. \\
More adaptive product direction & Keep evidence, assumptions, decisions, and changing conditions connected. \\
Better Product Management--R\&D collaboration & Make product rationale, technical constraints, and uncertainty more visible. \\
End-to-end product learning & Connect opportunities and decisions with realization and observed outcomes. \\
\bottomrule
\end{tabularx}
\end{table}

\section{The AI-Native Product Management Operating Model}
\label{sec:operating}

The previous section defined the vision and its principles. This section makes it tangible through an operating model, showing how humans and AI can work together while Product Management and R\&D remain closely connected.

The model uses the same basic vocabulary as recent work on AI-native R\&D \cite{britto2026ainative}. The remainder of the section introduces the main elements.

\subsection{Humans, AI Personas, Agents, Skills, and Shared Context}

The AI-native R\&D vision distinguishes AI personas\footnote{Here, an AI persona represents a professional perspective applied by AI, not a customer persona used in product discovery. It guides the questions, evidence, priorities, and trade-offs considered by AI.}, agents, and skills. A persona provides a professional perspective. An agent performs work within a defined scope. A skill provides a reusable capability. We adopt the same distinction for product management.

We propose six Product Management AI personas based on the scope of SPM\footnote{The personas specified here organize relevant professional perspectives rather than define a fixed and exhaustive role taxonomy} \cite{vandeweerd2006reference,ebert2007impact,maglyas2017core}. :
\begin{itemize}
    \item \textbf{Market Analyst.} Monitors customer evidence, competitors, market signals, ecosystems, and emerging trends.
    \item \textbf{Product Strategist.} Examines strategic fit, product direction, portfolio implications, and alternative scenarios.
    \item \textbf{Product Designer.} Explores customer problems, assumptions, product concepts, prototypes, and user experience.
    \item \textbf{Commercial Manager.} Examines pricing, business models, commitments, and contractual implications.
    \item \textbf{Lifecycle Manager.} Examines adoption, migration, realized value, service implications, and lifecycle choices.
    \item \textbf{Product Assurance Specialist.} Examines policies, regulations, risks, privacy, security obligations, and other assurance needs.
\end{itemize}

Several personas may contribute to one product question, while others may not be needed. Technical perspectives are not duplicated on the Product Management side. When technical expertise becomes relevant, R\&D professionals and the corresponding R\&D AI personas enter the same product-learning process.

A persona does not imply a dedicated agent. One agent may apply several personas, while specialist agents may be useful when knowledge, permissions, or methods differ. Agents perform work through reusable skills such as monitoring competitors, synthesizing evidence, comparing alternatives, generating scenarios, preparing prototypes, or identifying contradictory evidence.

A persona provides the professional lens, the agent executes the work, and skills provide reusable capabilities. They do not determine how responsibility is divided between humans and AI. That division depends on the activity and its consequences. Hence, we distinguish three modes of human--AI participation: 

\begin{itemize}
    \item \emph{AI-operated} activities are performed by AI within boundaries established by humans. Continuous monitoring, synthesis, and tracing changes are typical examples. 
    \item \emph{Human-directed} activities depend on human framing, judgment, or commitment, while AI provides evidence and analysis. Strategic choices and consequential product decisions belong to this category. 
    \item \emph{Collaborative} activities involve iterative contributions from humans and AI, such as exploring alternatives, prototyping, assessing feasibility, and interpreting outcomes.
\end{itemize}

The modes apply to activities rather than entire stages. Sensing can combine AI-operated monitoring with human-directed interpretation. Exploration is often collaborative. Consequential commitments remain human-directed. Realization and learning can combine all three modes as different professionals, personas, and skills become relevant.

Persistent participation also depends on shared product knowledge. If evidence, assumptions, decisions, and technical consequences must be reconstructed for every interaction, AI remains largely task-oriented. To address this aspect, we use a lightweight semantic layer as shared memory for humans and AI. It connects customers, market signals, competitors, opportunities, assumptions, decisions, product direction, technical capabilities, releases, and outcomes. Existing systems remain authoritative. The semantic layer preserves meaning, relationships, provenance, and decision rationale across them \cite{song2013semantic,xiao2018obda}.

As with AI-native R\&D, context is not consumed only by AI, and it evolves as work progresses. New evidence can update assumptions, technical findings can challenge product direction, and outcomes can be traced back to earlier decisions. The semantic layer helps humans and AI work from a consistent and traceable product context without creating another system of record.

Shared context does not imply unrestricted access. Security, privacy, and information boundaries must remain enforceable across the underlying systems and AI interactions.

\section{Illustrative Instantiation: A 5G Core Network Opportunity}
\label{sec:illustration}
\begin{figure}[H]
\centering
\includegraphics[width=\textwidth]{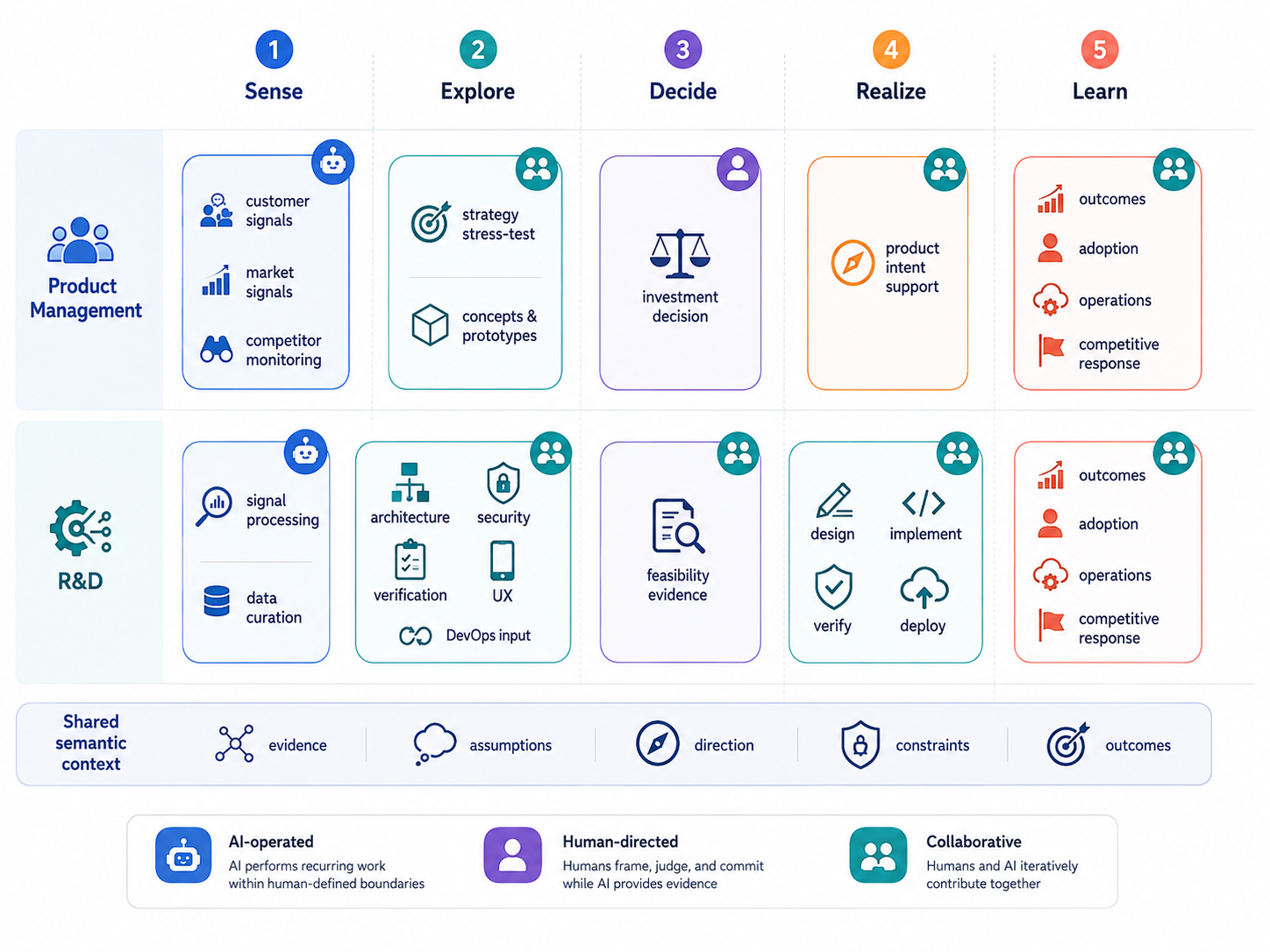}
\caption{Interleaved Product Management and R\&D participation in the illustrative 5G core case.}
\label{fig:operating-model}
\end{figure}

The case is illustrative rather than empirical. Its purpose is to show how the proposed elements can interact in a realistic product-management setting, not to validate their effects. Figure~\ref{fig:operating-model} presents a hypothetical 5G core network opportunity through the five activities of the product-learning loop. It shows Product Management and R\&D as interleaved participants and indicates whether representative activities are AI-operated, human-directed, or collaborative. The allocation is illustrative rather than fixed.

Several operators report difficulties assuring service levels for network slices used by enterprise customers. Their requested solutions differ. Support cases and product usage indicate related problems. These customer signals provide the starting point for sensing.

The Market Analyst AI persona also follows competitor offerings, operator strategies, technology developments, industry trends, and regulatory changes. This monitoring is largely AI-operated. AI connects the sources and surfaces a possible broader opportunity around slice-level service assurance.

A product manager interprets the evidence, its provenance, and remaining uncertainty. This activity is human-directed. The question is whether the combined signals represent an isolated customer issue, a broader change in operator needs, or an opportunity to differentiate the product.

Exploration becomes collaborative. The Product Strategist persona can compare competitors, examine plausible market developments, and generate alternative strategic responses. The product manager can challenge these alternatives and reframe the opportunity when the evidence suggests another interpretation.

The Product Designer persona separates the underlying problem from the requested solutions. One operator may ask for dashboards, another for stronger policy controls, and a third for automated remediation. AI can generate alternative concepts and prototypes. Human product professionals decide which assumptions deserve testing and interpret the resulting operator feedback.

As Figure~\ref{fig:operating-model} shows, R\&D enters before realization. Questions about telemetry, policy control, performance, architecture, security, and serviceability require technical expertise. Relevant R\&D professionals and AI personas join the exploration. Depending on the problem, these may include the architecture, security, verification, UX, and DevOps perspectives described in the AI-native R\&D vision \cite{britto2026ainative}.

This work is collaborative. AI personas can expose dependencies, compare technical alternatives, and perform bounded analysis. R\&D professionals interpret their implications and exercise engineering judgment. A technically simple dashboard may, for example, fail to address the underlying control problem. R\&D may also expose an alternative enabled by capabilities already under development.

The Product Strategist can then revisit the alternatives. Commercial and Lifecycle perspectives can be used to examine customer commitments, differentiation, deployment, and migration. AI connects the evidence and exposes uncertainty rather than reducing perspectives to a single score.

The investment decision remains human-directed. AI can prepare and challenge the evidence, but the human product authority decides whether to proceed and accepts responsibility for the resulting direction.

If the organization proceeds, R\&D receives more than a feature request. Shared context includes the customer and market evidence, strategic rationale, expected outcome, alternatives, assumptions, prototype findings, technical constraints, and unresolved questions. Product intent remains visible during realization.

Realization can contain all three participation modes. Some activities may be AI-operated within defined boundaries. Others require collaboration between R\&D professionals and AI personas. Consequential technical choices remain subject to appropriate human judgment.

Product Management also remains engaged. An architecture dependency may change cost or timing. Verification may weaken an expected outcome. A technical development may create an option that was unavailable during the original exploration. Persistent AI connects these changes back to the product reasoning.

Learning closes the case without ending it. AI can continuously connect operator feedback, adoption, competitor responses, commercial results, verification evidence, and operational behavior with the assumptions behind the decision. Product and R\&D professionals interpret what these outcomes mean for subsequent direction.

The example highlights three characteristics of the operating model. First, responsibility does not shift uniformly from humans to AI. Activities can be AI-operated, human-directed, or collaborative, depending on the judgment and consequences involved.

Second, Product Management and R\&D do not interact through a fixed handoff. R\&D expertise can enter the exploration phase, while Product Management remains connected throughout realization and learning.

Third, persistent AI provides continuity across these interactions. Product choices remain connected to customer and market evidence, strategic reasoning, and expected outcomes. Technical assessments remain connected to dependencies, alternatives, uncertainty, and later outcomes.

This continuity does not remove disagreement. Product professionals should challenge technical assumptions, and R\&D should challenge product assumptions. The expected benefit is better-informed disagreement based on greater visibility into the reasoning on both sides.

We expect the combined model to improve strategic exploration, mutual understanding, and end-to-end ownership. These are propositions that require empirical evaluation.

\section{Research Agenda}
\label{sec:research-agenda}

While the vision outlined here indicates how AI can improve product management and its relationship with AI-native R\&D, there are still aspects that require further exploration. For example, AI-native does not imply autonomous product management, and the boundary is an open question. Furthermore, AI-generated information also requires interpretation, since raw use can amplify weak evidence or create artificial certainty \cite{leonardi2026certainty}.

The principle of challenge before consensus also requires care. Multiple AI personas may appear to broaden the analysis while producing similar reasoning when they rely on the same models or context. In story-writing experiments, AI support improved individual creativity but reduced collective diversity \cite{doshi2024creativity}, while deliberately varied personas helped preserve diversity \cite{wan2026personas}. These findings motivate caution, not a conclusion about product management. Persona diversity should not be treated as evidence of independent reasoning. Product professionals must still examine sources, seek counter-evidence, and challenge shared assumptions.

Finally, persistent AI can connect knowledge across products and functions, but errors in context can also propagate further. Large-scale settings need stronger shared context, specialization, and coordination without changing the underlying vision.

We have consolidated aspects requiring further exploration into four questions, which are detailed below.

\subsection{Persistent AI Participation}

How does persistent AI participation change product management compared with task-level assistance? Current research remains largely task-oriented \cite{witkowski2025ai,moriserra2026llm}. Studies should examine what teams notice, how quickly they respond, the quality of product decisions, and the effort required to coordinate product work.

\subsection{AI-Supported Exploration}

When does AI broaden product exploration, and when does it reinforce existing patterns? AI can identify opportunities, compare alternatives, and challenge assumptions, but the inferences it generates should not be treated as evidence. Research should examine whether AI expands the opportunity space or mainly produces variations of familiar options \cite{narver2004responsive,day1994}.

\subsection{Human--AI Allocation and Product Judgment}

Which product-management activities should be AI-operated, human-directed, or collaborative? The appropriate mode may depend on uncertainty, reversibility, consequences, and organizational context \cite{parasuraman2000automation,fuegener2022delegation}. A related question is how sustained use of AI affects human judgment. Research should examine how junior and senior professionals develop, maintain, and calibrate judgment when AI participates regularly in their work, including when overreliance emerges \cite{bucinca2021trust}.

\subsection{AI Across the Product Management--R\&D Boundary}

How does shared AI participation affect coordination between Product Management and R\&D? Shared context may make product and technical reasoning more visible, but it may also cause both functions to rely on the same incorrect interpretation. Research should examine whether shared AI participation improves constructive disagreement, continuity from opportunity to outcome, and coordination across the two functions.

\section{Conclusion}
\label{sec:conclusion}

AI-native product management represents a shift from task-level assistance toward persistent AI participation in product learning. Human professionals remain responsible for consequential product judgment and accountability.

The connection with AI-native R\&D is central. Product management can strengthen the ability to determine and reconsider what deserves realization, while AI-native R\&D may increase the capacity to realize it. Persistent AI can connect product reasoning with technical possibilities, constraints, and outcomes.

The vision remains conceptual, and we acknowledge that an empirical evaluation is necessary to establish its effectiveness and measure its impact. We also outlined areas that require further exploration to enable further evolution of the vision described in this paper.

\bibliographystyle{splncs04}
\bibliography{references}

\end{document}